\documentclass[11pt,a4paper]{article}

\usepackage[T1]{fontenc}
\usepackage{lmodern}
\usepackage[margin=1in]{geometry}
\usepackage{amsmath}
\usepackage{graphicx}
\usepackage{siunitx}
\usepackage[numbers,square,sort&compress]{natbib}
\usepackage[hidelinks]{hyperref}

\title{Randomly Coupled Multicore Fibers for Data-Efficient Speckle Sensing}
\author{Mingjing Xu, Lin Ma\thanks{Corresponding author. \href{mailto:ma.lin@sjtu.edu.cn}{ma.lin@sjtu.edu.cn}.}, Tianqi Cheng, Fengrui Yu, and Zuyuan He\\
\small State Key Laboratory of Photonics and Communications\\
\small Shanghai Jiao Tong University, Shanghai 200240, China}
\date{}

\begin{document}
\maketitle

\begin{abstract}
Mode coupling in optical fibers enables external perturbations to be encoded into spatial intensity distributions, providing an effective basis for optical sensing. However, quantitative speckle sensing often relies on numerous pixel-level features for signal demodulation. Here, randomly coupled multicore fibers are demonstrated as a structured sensing platform in which random inter-core coupling encodes bending-induced perturbations into core-resolved intensity distributions. The fixed multicore geometry provides natural sampling channels, yielding only four or seven normalized core-intensity features per image. Compared with representative machine-learning-assisted fiber sensors, this physically structured readout reduces the number of scalar model inputs per record by several-fold to more than four orders of magnitude. The output intensity distribution remains stable over 12~h under unchanged conditions and responds more strongly to bending than to axial stretching or twisting in the tested configurations. For the four-core fiber, the multilayer perceptron and gradient-boosted regression trees achieve curvature root-mean-square errors of $0.014~\mathrm{m}^{-1}$ and $0.022~\mathrm{m}^{-1}$, respectively, on curvature states held out from training. These results demonstrate the potential of random inter-core coupling for compact and data-efficient optical fiber sensing.
\end{abstract}


\section{Introduction}
Optical fiber speckles arise from the interference of multiple coherent modes and encode propagation information into complex spatial intensity patterns with rich information content~\cite{ploschner2015seeing,tzang2018adaptive,li2021compressively}. Although speckle patterns appear random, their spatial variations encode information about the incident field, the propagation medium and external perturbations. The key challenge is to recover this information from complex and perturbation-sensitive intensity patterns. Recent advances in transmission-matrix characterization~\cite{kupianskyi2024untangling}, wavefront shaping~\cite{han2023tailoring} and deep-learning-assisted reconstruction~\cite{rahmani2018multimode} have shown that speckle fields can be decoded, controlled and engineered for specific optical functions. These capabilities have enabled broad applications in imaging~\cite{stellinga2021time,luo2022computational}, spectroscopy~\cite{xu2024wgm}, optical communications~\cite{gao2025semantic} and quantum information processing~\cite{leedumrongwatthanakun2020programmable}. For sensing, the intrinsic sensitivity of modal interference to external perturbations allows environmental information to be encoded in speckle evolution, making optical speckles valuable resources for fiber sensing~\cite{facchin2024intrinsic}.

Among fiber speckle sensing platforms, multimode fibers (MMFs) have been widely investigated~\cite{chapalo2024review}. Their strong sensitivity to external perturbations, however, is accompanied by complex and nonlinear speckle evolution, which complicates direct demodulation~\cite{smith2022specklegram}. Early MMF schemes mainly used correlation analysis to infer perturbations from global changes in speckle-pattern similarity~\cite{zhou2018diffusing,huang2018dna}. Because faster decorrelation can increase sensitivity while shortening the unambiguous measurement range, these methods commonly face a sensitivity--range trade-off. Machine learning enables quantitative sensing by learning the nonlinear relation between speckle evolution and physical quantities directly from data~\cite{bielak2026polarization}, with demonstrations in biological~\cite{zhou2021functional}, mechanical~\cite{tao2026nanosecond} and physical-field sensing~\cite{xiong2024polarimeter}. Most learning-assisted MMF schemes, however, extract many pixel-level intensity values from selected regions of two-dimensional speckle images, which increases input dimensionality and can increase model and training-data requirements~\cite{newaz2023review}. Feature extraction reduces the number of model inputs but discards part of the measured speckle information~\cite{xia2024nonlinear}. Fibers supporting fewer or more stable modes can generate smoother and more reproducible speckle evolution, although reduced modal complexity may also weaken the perturbation response~\cite{li2024ultrashort,wei2021limited}. A sensing platform that combines few measured features with stable and perturbation-sensitive speckle evolution is therefore desirable for accurate and data-efficient regression~\cite{wei2021limited,fu2024onn}.

Multicore fibers (MCFs) support multiple spatially fixed optical channels within a single fiber~\cite{vandenhout2025pinnacle,richardson2013sdm} and have been widely explored for imaging~\cite{sun2022qpi,zolnacz2024endoscopic}, fiber lasers~\cite{klenke2018laser}, optical communications~\cite{li2026hybrid}, quantum information processing~\cite{zahidy2024qkd} and sensing~\cite{caucheteur2022sensors,wolf2022fbg,matsui2022wcmcf}. Compared with single-mode fibers (SMFs) and MMFs, MCFs provide a structured propagation geometry in which the optical field is distributed over a finite number of discrete cores rather than a continuously distributed output intensity pattern. Inter-core coupling, an intrinsic property of closely spaced MCFs, is generally regarded as crosstalk that affects transmission performance, and the coupling condition is strongly affected by fiber bending~\cite{diamandi2017crosstalk,soeiro2018bending,sakamoto2024nonlinear}. Our previous work showed that inter-core coupling in randomly coupled multicore fibers (RC-MCFs) can instead be exploited for ultrahigh-sensitivity curvature sensing through coupling-induced spectral evolution~\cite{xu2026lpr}. However, spectral approaches~\cite{hu2025lstm,li2025tactile} rely on wavelength-resolved interrogation and do not directly exploit the spatially discrete output channels inherent to the multicore architecture. In an RC-MCF, the core positions are fixed by the fiber geometry, whereas the intensity distribution across the cores evolves through perturbation-dependent inter-core energy redistribution~\cite{arrizabalaga2020vector}. The coexistence of fixed spatial channels, random inter-core coupling and perturbation sensitivity defines an intermediate structured regime between SMF stability and MMF speckle complexity, with the potential for stable sensing from a small number of measured intensity features.

This work presents a data-efficient fiber speckle sensing scheme based on RC-MCFs as an alternative to pixel-intensive MMF readout. The fixed multicore geometry samples coupling-induced speckle evolution through only a few normalized core intensities per image, rather than through many pixel-level intensity values from a selected speckle-image region. Random inter-core coupling, commonly regarded as crosstalk in MCF transmission systems, is thereby converted into a physical encoding resource for perturbation-sensitive sensing. The RC-MCF output combines high temporal stability without intentional perturbations with a curvature-selective response under mechanical deformation. Its core-intensity patterns also evolve continuously across different operating wavelengths. Lightweight machine-learning models, including gradient-boosted regression trees (GBRT) and a multilayer perceptron (MLP), enable high-accuracy curvature reconstruction from the resulting core-intensity vectors without requiring complex image-based neural networks. A cross-study comparison further evaluates the reduction in per-record input requirements relative to representative image-, spectrum- and voltage-based fiber sensors. These results establish RC-MCFs as stable, compact and data-efficient sensing platforms that reduce the number of intensity features required for fiber speckle demodulation.

\section{Results}

\subsection{Sensing principle and experimental implementation}

The RC-MCFs comprise closely spaced step-index cores within a common cladding. The tested fibers had a core diameter of 9~$\mu$m, a cladding diameter of 125~$\mu$m, and a relative refractive-index difference of 0.37\% between the core and cladding. The core pitches ranged from 15 to 25~$\mu$m, depending on the sample~\cite{xiong2024corepitch}. 

\begin{figure}[!htb]
    \centering
    \includegraphics[width=\linewidth]{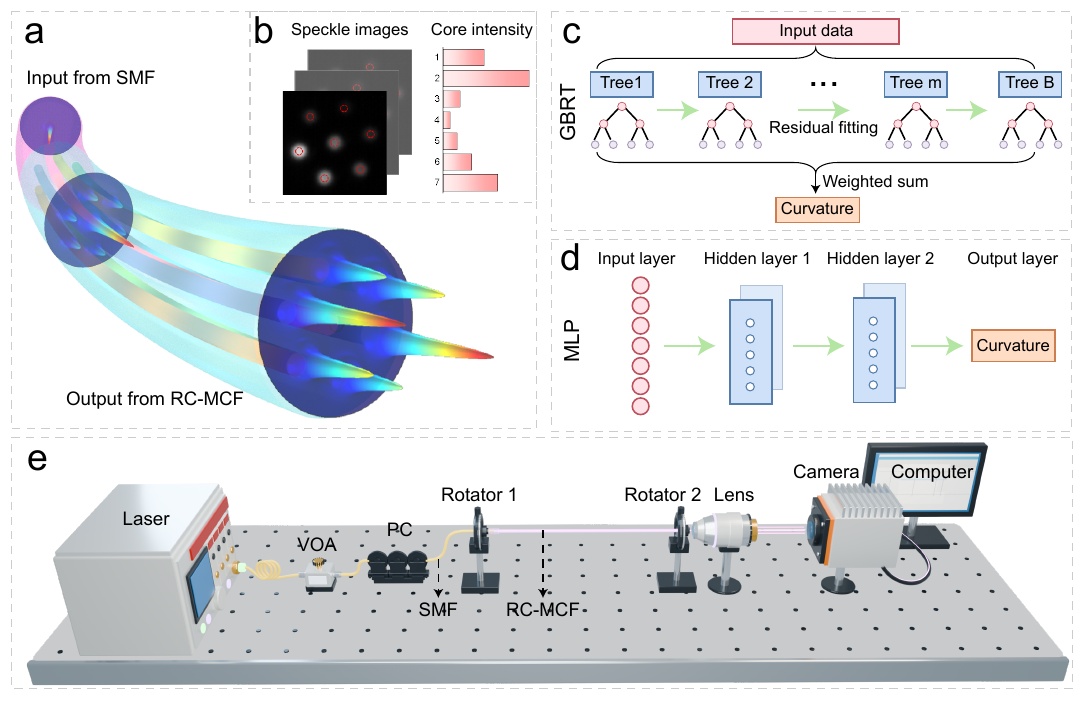}
    \caption{{RC-MCF speckle sensing system with data-efficient readout.}
    \textbf{a}, Schematic illustration of light propagation and
    inter-core coupling in the RC-MCF.
    \textbf{b}, Output speckle images and extracted core intensity features.
    \textbf{c}, GBRT model for curvature reconstruction using
    ensemble decision trees.
    \textbf{d}, MLP model for curvature regression based on
    the extracted core-intensity vector.
    \textbf{e}, Experimental setup of the RC-MCF speckle sensing system.}
    \label{fig:fig1}
\end{figure}

When each core supports a single guided mode, the evolution of the complex modal amplitude $a_i(z)$ in core $i$ can be described by Equation~\eqref{eq:intercore-coupling}~\cite{xu2026lpr}.
\begin{equation}
    \frac{\mathrm{d}a_i(z)}{\mathrm{d}z}
    = -\mathrm{j}\beta_i(z)a_i(z)
      -\mathrm{j}\sum_{k\ne i}\kappa_{ik}(z)a_k(z).
    \label{eq:intercore-coupling}
\end{equation}
Here, $z$ is the propagation coordinate, $\beta_i(z)$ is the local propagation constant of core $i$, and $\kappa_{ik}(z)$ is the local coupling coefficient from core $k$ to core $i$. Longitudinal variations in the propagation constants and coupling coefficients give rise to random inter-core coupling and complex evolution of the optical field among the cores. For a given launch condition and bending orientation, changing the curvature modifies the propagation-constant mismatches $\Delta\beta_{ik}(z)=\beta_i(z)-\beta_k(z)$ and hence the phase matching between cores. Bending can also modify $\kappa_{ik}(z)$ through changes in modal overlap. The resulting power redistribution changes the output power in each core, which scales as $|a_i(L)|^2$ at the end of the sensing fiber of length $L$. The curvature-dependent output is read out through the spatially resolved core intensities, as illustrated in Figure~\ref{fig:fig1}(a).

Figure~\ref{fig:fig1}(b) illustrates the extraction of core-intensity
features from the camera images. Core positions were identified from
the resolved output spots and the transverse core arrangement.
Circular regions of interest (ROIs) with a radius of 3 pixels were
manually centered on the spots in the first image and kept fixed
throughout each measurement sequence. The mean grayscale intensity $I_i$ within the ROI of core $i$ was used as a proxy for its output power. The intensities were normalized according to
Equation~\eqref{eq:normalized-core-intensity}.
\begin{equation}
    \begin{aligned}
        p_i &= \frac{I_i}{\sum_{k=1}^{N_{\mathrm{c}}} I_k},\\
        \mathbf{p} &= [p_1,\ldots,p_{N_{\mathrm{c}}}]^{\mathrm{T}},
    \end{aligned}
    \label{eq:normalized-core-intensity}
\end{equation}
where $N_{\mathrm{c}}$ is the number of detected cores. The complete
vector $\mathbf{p}$ was used as the input to the gradient-boosted
regression trees (GBRT) and multilayer perceptron (MLP) models shown in
Figure~\ref{fig:fig1}(c,d). The four-core fiber (4CF) and seven-core fiber (7CF) thus provided four and seven
input features per image, respectively. Because
$\sum_{i=1}^{N_{\mathrm{c}}}p_i=1$, these vectors have at most three and
six independent degrees of freedom, respectively.

The experimental setup used to implement the RC-MCF speckle sensing scheme is shown in Figure~\ref{fig:fig1}(e). Light from a tunable laser was coupled into an SMF, with its optical power and polarization state adjusted using a variable optical attenuator (VOA) and a polarization controller (PC), respectively. The SMF was fusion-spliced to the RC-MCF using a Fujikura FSM-100P+ fusion splicer, with the SMF core aligned to the selected input core of the RC-MCF. Fiber deformation was controlled using translation and rotation stages. The output field was collected using a $20\times$ objective lens with a numerical aperture of 0.45 and imaged onto an infrared camera (Bobcat 320, Xenics). The exposure time was 1~ms, and the electronic gain was kept constant throughout each measurement sequence.

\subsection{Coupling and speckle characteristics of RC-MCFs}

A measurable sensing response in an RC-MCF requires sufficient inter-core coupling over the test length, whereas the data-efficient readout used here benefits from fundamental-mode-dominated output in each core. The inter-core coupling state depends on the operating wavelength, propagation length, core pitch, and transverse core arrangement. We therefore measured the output intensity patterns of several RC-MCF samples at different wavelengths and propagation lengths to characterize their inter-core coupling and intra-core modal states.

\begin{figure}[tbp]
    \centering
    \includegraphics[width=\linewidth]{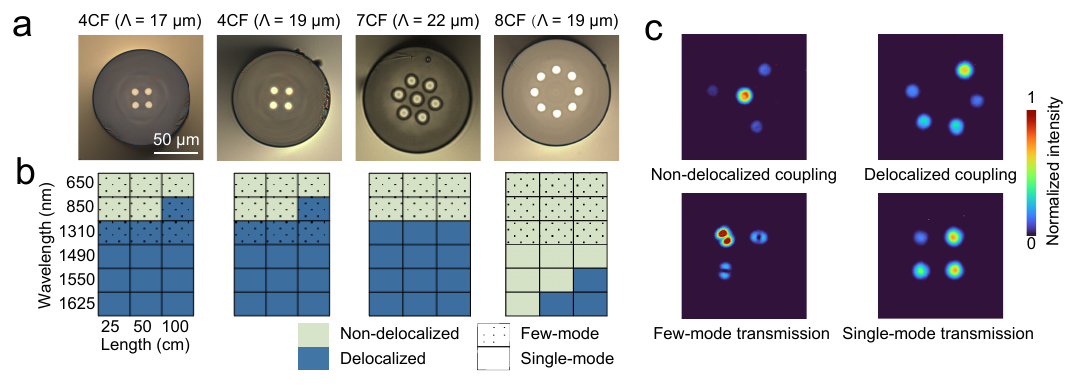}
    \caption{{Coupling and speckle characteristics of RC-MCFs.}
        \textbf{a}, Cross-sectional micrographs of the tested 4CFs, 7CF and 8CF samples.
        \textbf{b}, Coupling and modal transmission states versus wavelength and fiber length.
        \textbf{c}, Output patterns of non-delocalized and delocalized 7CF and few-mode and single-mode 4CF. All intensity patterns are independently normalized.}
    \label{fig:fig2}
\end{figure}

Figure~\ref{fig:fig2}(a) shows cross-sectional micrographs of four RC-MCF samples: two square-array 4CFs with core pitches of 17 and 19~$\mu$m, a 7CF with a central core surrounded by six cores at a pitch of 22~$\mu$m, and a single-ring eight-core fiber (8CF) with a pitch of 19~$\mu$m. Figure~\ref{fig:fig2}(b) maps the coupling states observed at different wavelengths and propagation lengths. Small deviations from the nominal fiber geometry can affect both the coupling length and the modal cutoff wavelength, so the regimes were identified from the measured output patterns. In the non-delocalized regime, the launched core remains dominant even
when some power transfers to neighboring cores. In the delocalized regime, optical power is distributed across multiple cores. The upper row of Figure~\ref{fig:fig2}(c) shows representative 7CF patterns in both regimes. Samples operating in the delocalized regime were selected for sensing so that perturbation-induced intensity changes could be measured across multiple cores.

The modal state within each core also affects the reliability of intensity extraction. At shorter wavelengths, individual cores may support few-mode transmission, producing asymmetric or spatially structured intra-core intensity patterns, as shown in the lower-left image of Figure~\ref{fig:fig2}(c). The readout method benefits from the approximately Gaussian and centrosymmetric output produced under single-mode transmission, whereas few-mode output can cause the sampled intensity to depend on the intra-core modal distribution rather than solely on the coupling state. Considering both the inter-core coupling state and the modal transmission condition, the 4CF with $\Lambda = 19~\mu$m and the 7CF with $\Lambda = 22~\mu$m  were selected for the subsequent speckle transmission and sensing experiments within their respective single-mode operating wavelength ranges.

\begin{figure}[!htb]
    \centering
    \includegraphics[width=\linewidth]{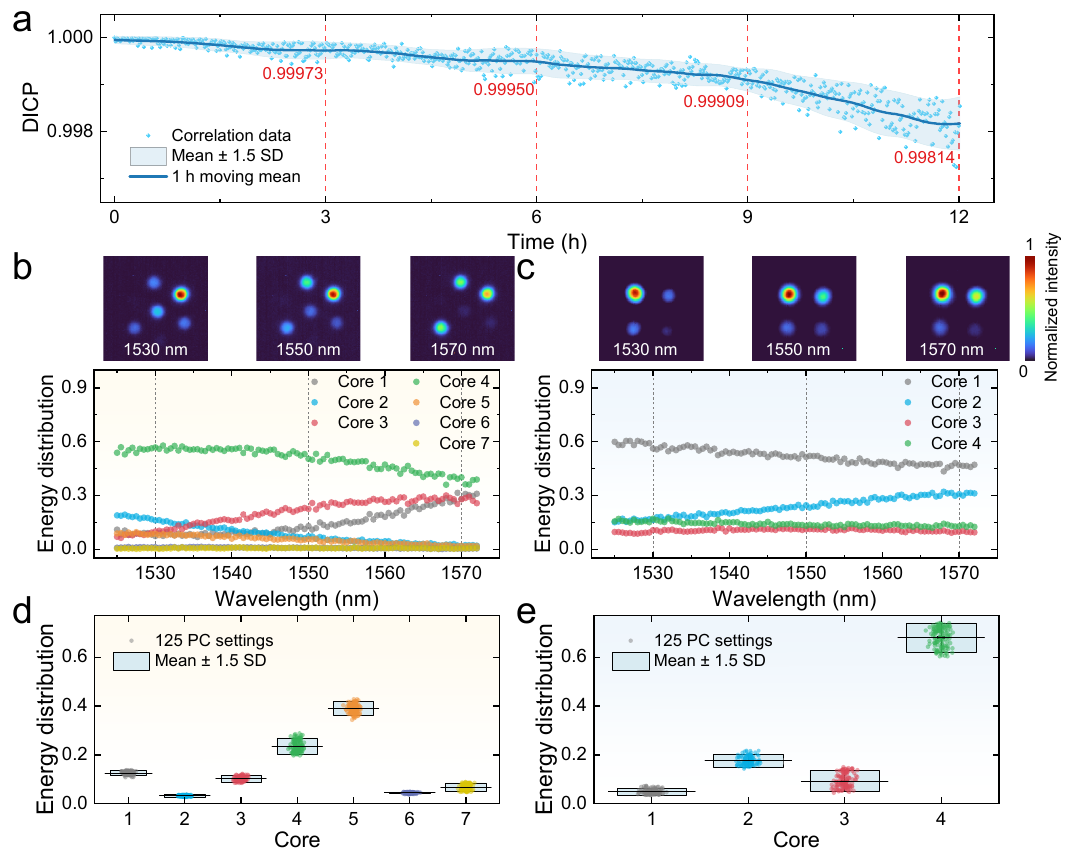}
    \caption{Temporal stability, wavelength-dependent and polarization-dependent speckle evolution of RC-MCFs.
    \textbf{a}, Long-term stability of the 7CF.
    \textbf{b, c}, Wavelength-dependent normalized core intensities of the 7CF and 4CF, respectively, with representative output patterns at 1530, 1550 and 1570~nm.
    \textbf{d, e}, Normalized core-intensity distributions of the 7CF and 4CF, respectively, measured across 125 polarization-controller configurations.}
    \label{fig:fig3}
\end{figure}

Temporal stability of the output speckle pattern is essential for reliable RC-MCF sensing. A 20-cm-long 7CF with a core pitch of $\Lambda = 22~\mu$m was monitored at 1550~nm under ambient laboratory conditions, with output images acquired every 1~min over 12~h. The degree of intensity correlation of patterns (DICP) between the initial reference image $I_0(x,y)$ and the $q$th recorded image $I_q(x,y)$ was calculated as
\begin{equation}
    \mathrm{DICP}(q) =
    \frac{
    \sum_{x,y}
    \left[I_0(x,y) - \overline{I_0}\right]
    \left[I_q(x,y) - \overline{I_q}\right]
    }{
    \sqrt{
    \sum_{x,y}\left[I_0(x,y) - \overline{I_0}\right]^2
    \sum_{x,y}\left[I_q(x,y) - \overline{I_q}\right]^2
    }},
    \label{eq:dicp}
\end{equation}
where $q$ is the image index, and the sums and mean intensities $\overline{I_0}$ and $\overline{I_q}$ were evaluated within the predefined core regions. These regions were confined within the illuminated core spots, and no separate background subtraction was applied. A DICP value close to 1 indicates high correlation with the initial pattern. As shown in Figure~\ref{fig:fig3}(a), the DICP decreased only slightly over time. Its 1-h moving mean remained above 0.999 during the first 9~h and above 0.998 throughout the 12-h measurement, indicating high temporal stability of the measured core-region intensity pattern. The shaded region represents the moving mean $\pm$ 1.5 standard deviations.

The wavelength and polarization dependence of the output pattern was subsequently examined. Figure~\ref{fig:fig3}(b,c) shows the normalized core-intensity distributions of the 7CF with $\Lambda = 22~\mu$m and the 4CF with $\Lambda = 19~\mu$m, respectively. During wavelength tuning from 1525 to 1570~nm in steps of 0.5~nm, the normalized core intensities generally evolve continuously, with only minor local fluctuations. Representative output patterns at 1530, 1550 and 1570~nm further illustrate this gradual evolution. These results show that wavelength-induced changes in inter-core coupling are reflected in a smooth and traceable output response. The polarization dependence was further evaluated using 125 polarization-controller configurations, obtained by independently setting the three paddles from $0^\circ$ to $180^\circ$ in $45^\circ$ increments. As shown in Figure~\ref{fig:fig3}(d,e), although the individual core intensities vary with the polarization-controller setting, the overall core-intensity distribution remains relatively consistent, indicating moderate polarization dependence and good robustness under the tested input polarization conditions.

\subsection{Sensing characteristics of RC-MCFs}

\begin{figure}[tbp]
    \centering
    \includegraphics[width=\linewidth]{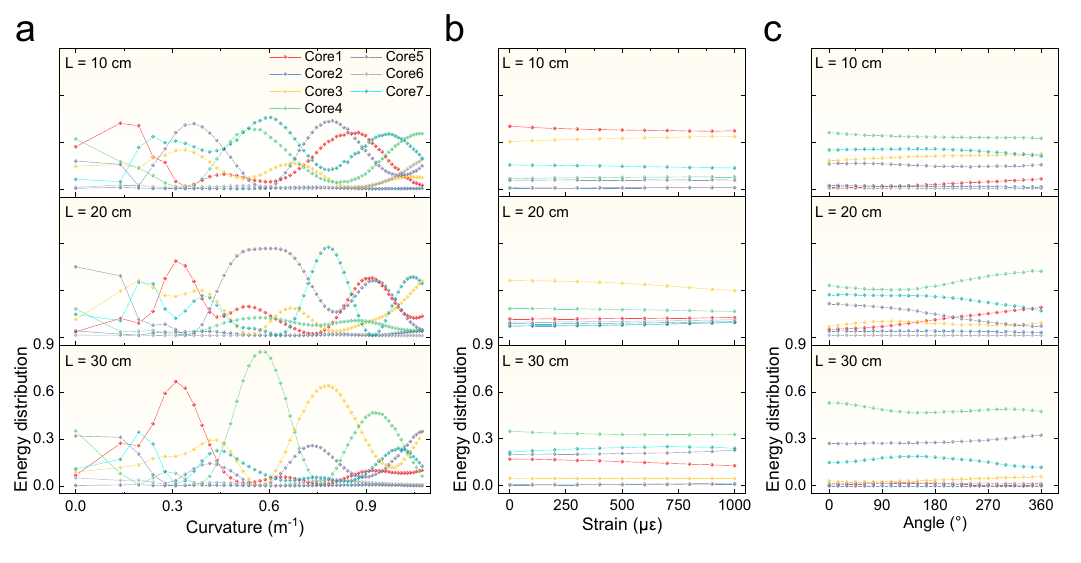}
    \caption{{Core-intensity responses of 7CFs to mechanical deformation. Normalized core intensities were measured at 1550 nm for fiber lengths of 10, 20, and 30~cm.}
        \textbf{a}, Bending.
        \textbf{b}, Axial stretching.
        \textbf{c}, Twisting.}
    \label{fig:fig4}
\end{figure}

Changes in the inter-core coupling state are reflected in the output core-intensity distribution. Figure~\ref{fig:fig4} summarizes the bending, axial-stretching and twisting responses of three 7CF samples with propagation lengths of 10, 20 and 30~cm tested at 1550~nm. Deformations were applied using translation and rotation stages. The initial effective clamping length was 50~cm for all measurements, independently of the RC-MCF propagation length.

For bending, the clamp separation was decreased in 0.1~mm steps over a total displacement of 6~mm. Curvature was defined as $C = 1/R$, where $R$ is the bending radius. 61 curvature values were calculated from the clamp displacement using a circular-arc approximation and represent the nominal curvature of the prescribed bending configuration. The resulting range was 0--1.07~m$^{-1}$, with equal displacement steps corresponding to nonuniform curvature intervals. Figure~\ref{fig:fig4}(a) shows pronounced, continuous oscillatory redistribution of core intensities for all three samples. Within this gentle-bending regime, with nonzero curvatures corresponding to meter-scale bending radii, successive small curvature increments produced measurable changes in the output patterns.

For axial stretching, the clamp separation was increased in 0.05~mm steps over 0.5~mm, corresponding to $0$--$1000~\mu\varepsilon$. Figure~\ref{fig:fig4}(b) shows only minor changes in normalized core intensities. For twisting, one clamp was rotated from $0^\circ$ to $360^\circ$ in $15^\circ$ steps, reaching a maximum twist rate of approximately 12.6~rad~m$^{-1}$. As shown in Figure~\ref{fig:fig4}(c), twisting induced slightly greater intensity variations than axial stretching, particularly in longer samples, but substantially weaker variations than bending.

For comparison of deformation scales, the bending-induced axial strain at core $i$ was estimated as $|\varepsilon_i|\approx C|y_i|$, where $y_i$ is the perpendicular distance from the core center to the bending neutral axis. At the maximum curvature, the upper-bound strain at an outer-core center was approximately $24~\mu\varepsilon$, whereas the maximum twisting-induced outer-core shear strain was approximately $280~\mu\varepsilon$. These estimates depend on the core positions and, for bending, the orientation of the multicore geometry relative to the neutral axis, and serve only to compare the deformation scales.

\begin{figure}[tbp]
    \centering
    \includegraphics[width=\linewidth]{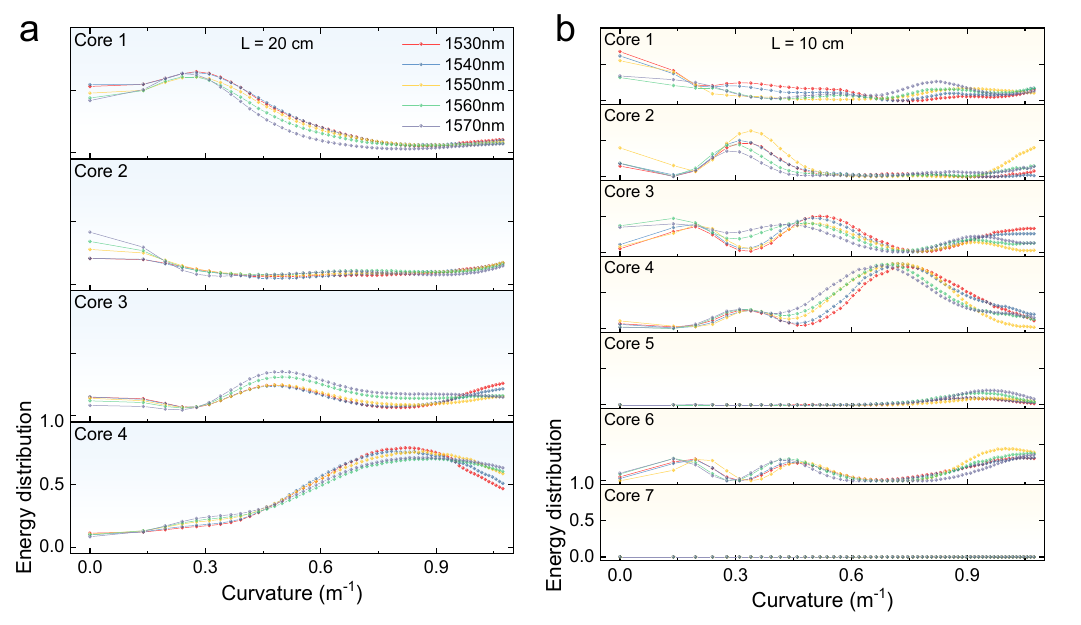}
    \caption{{Wavelength dependence of RC-MCF curvature sensing. Normalized core intensities versus curvature at wavelengths from 1530 to 1570~nm.}
        \textbf{a}, 20~cm 4CF.
        \textbf{b}, 10~cm 7CF.}
    \label{fig:fig5}
\end{figure}

This deformation selectivity arises from the different changes in the relative propagation constants of the cores. Axial stretching predominantly produces common-mode propagation-constant shifts, whereas twisting introduces only limited core-to-core propagation-constant differences under the tested conditions. Bending, however, generates a transverse strain gradient across the fiber cross-section, causing cores on opposite sides of the neutral axis to experience different or opposite propagation-constant shifts. The resulting change in the relative phase-matching condition produces substantial core-intensity redistribution. Although the detailed bending trajectories differ among the 10, 20 and 30~cm samples, their variation amplitudes and oscillation frequencies do not increase monotonically with fiber length. Propagation length therefore affects the accumulated coupling trajectory but does not provide a simple monotonic increase in bending response. These observations identify bending as the dominant deformation perturbation of the RC-MCF coupling state within the tested ranges and confirm that curvature is continuously encoded into the output intensity distribution.

The normalized core intensities of the 20-cm 4CF and the 10-cm 7CF are plotted against curvature in Figure~\ref{fig:fig5}(a,b), respectively. Both fibers exhibited systematic curvature-dependent responses at all five wavelengths tested from 1530 to 1570~nm in 10~nm steps. The trajectories shift gradually without abrupt changes between neighboring wavelengths, showing that the curvature response varied gradually across the five sampled wavelengths.

\subsection{Data-efficient curvature reconstruction}

Machine-learning regression was used to reconstruct curvature from the non-monotonic core-intensity responses. The models were trained and evaluated using bending datasets from the 10-cm 7CF at 1310~nm and the 20-cm 4CF at 1570~nm. Each image was represented by the normalized core-intensity vector $\mathbf{p}$ defined in Equation~\eqref{eq:normalized-core-intensity}. Two lightweight regression models were evaluated for the compact numerical features produced by the RC-MCF readout. For both models, each recorded image constituted one sample, pairing its normalized core-intensity vector with the corresponding curvature label. Additional images increased the number of samples rather than the dimension of an individual input vector.

GBRT constructs an additive ensemble by sequentially fitting regression trees to the residual errors of the preceding ensemble~\cite{bentejac2021comparative}. Its established performance on tabular data motivated its use for core-intensity regression~\cite{hollmann2025accurate}. The model was implemented using \textrm{fitrensemble} with least-squares boosting, 150 learning cycles, a learning rate of 0.05 and a maximum of 15 decision splits per tree. The MLP is a feedforward neural network trained through back-propagation~\cite{rumelhart1986backprop}. Related neural-network regression methods have been applied to spatially resolved MMF sensing~\cite{smith2024complex}. Here, the MLP learned the nonlinear mapping between core-intensity features and curvature. Implemented using \textrm{trainNetwork}, it comprised a feature input layer, two fully connected hidden layers with 32 neurons each and a scalar regression output. The hidden layers used hyperbolic tangent activation. The network parameters were optimized using the Adam algorithm~\cite{Kingma2015Adam} to minimize the training-set mean squared error, with an initial learning rate of $10^{-3}$, a mini-batch size of 32 and a maximum of 800 epochs. The intensity-normalized inputs were used without additional feature-wise standardization, and the training samples were shuffled once before training.

Prediction performance was evaluated using three data-partition strategies. As an auxiliary within-dataset baseline, all-random splitting randomly assigned individual images to the training and test sets with a ratio of 0.9:0.1. For the curvature evaluation, curvature-random splitting assigned complete curvature groups to the training and test sets using the same 0.9:0.1 ratio, such that all images acquired at a given curvature were confined to a single set. This strategy was used for the quantitative comparison with previously reported methods because it evaluates prediction at unseen curvature states while preventing images acquired at the same curvature state from appearing in both the training and test sets. Cross-measurement reproducibility was further assessed using repeated-test evaluation, in which the first curvature scan was used for training and the second scan for testing. The fiber was not remounted between the two scans in order to preserve the inter-core coupling state. Remounting can modify the initial output intensity distribution and the subsequent curvature-dependent intensity evolution, in which case model retraining would be required.

\begin{figure}[!htb]
    \centering
    \includegraphics[width=\linewidth]{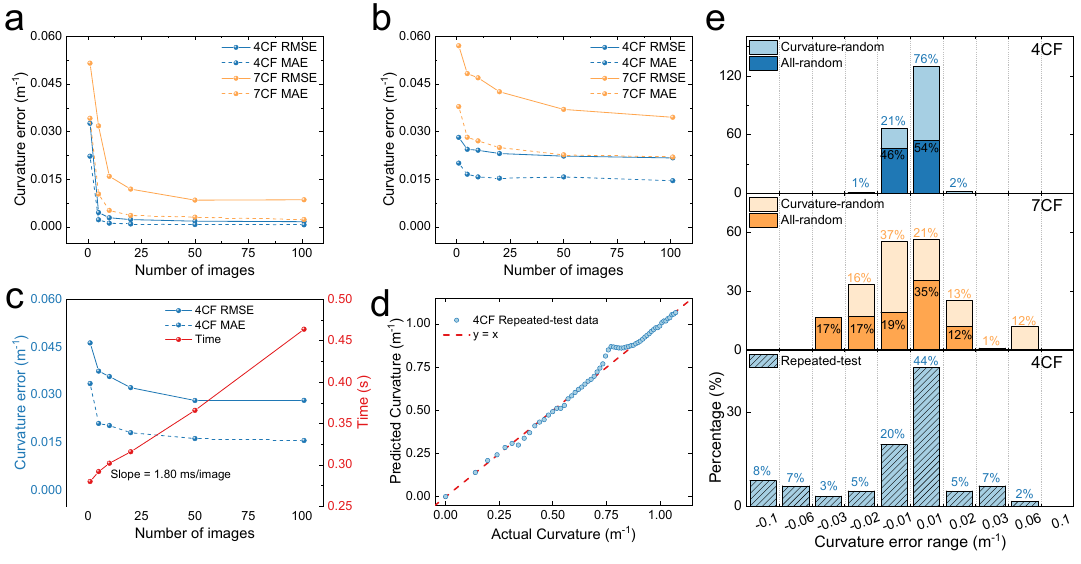}
    \caption{{GBRT-based curvature reconstruction using RC-MCF core-intensity features.}
        \textbf{a}, RMSE and MAE versus the number of images per curvature state for the 4CF and 7CF under all-random splitting.
        \textbf{b}, Corresponding errors under curvature-random splitting.
        \textbf{c}, RMSE, MAE, and total training and prediction time for the 4CF in the repeated-test evaluation.
        \textbf{d}, Predicted versus reference curvature for the second 4CF scan.
        \textbf{e}, Error distributions for the 4CF and 7CF under both random splitting strategies, and for the 4CF under repeated-test evaluation.The results in (a)--(c) are averaged over 200 evaluations, whereas (d) and (e) show representative individual evaluations.}
    \label{fig:fig6}
\end{figure}

Prediction performance was quantified using the root mean square error (RMSE) and mean absolute error (MAE),
\begin{equation}
    \mathrm{RMSE}
    =
    \sqrt{\frac{1}{M}\sum_{j=1}^{M}
    \left(\hat{C}_j-C_j\right)^2},
    \label{eq:rmse}
\end{equation}
and
\begin{equation}
    \mathrm{MAE}
    =
    \frac{1}{M}\sum_{j=1}^{M}
    \left|\hat{C}_j-C_j\right|,
    \label{eq:mae}
\end{equation}
where $C_j$ and $\hat{C}_j$ are the reference and predicted curvatures of the $j$th test sample, respectively, and $M$ is the number of test samples. Where random partitioning or model initialization was involved, training and evaluation were repeated 200 times, and quantitative comparisons used the averaged error metrics. These repetitions served statistical evaluation and are not required during practical sensing operation. The individual prediction-error distribution shown in Figure~\ref{fig:fig6}(e) was obtained from one illustrative evaluation run.

All models were implemented in MATLAB R2023a and evaluated on the same laptop equipped with an AMD Ryzen 7 5800H CPU, 16~GB RAM and an NVIDIA GeForce RTX 3060 Laptop GPU with 6~GB memory. Computation time included model training and prediction for one evaluation run but excluded data loading, ROI-based feature extraction and data partitioning. The fitted time slopes describe the increase in total training-and-prediction time per additional image per curvature state, rather than single-sample inference latency.

Figure~\ref{fig:fig6} summarizes the GBRT reconstruction results. Up to 101 images were acquired at each curvature state at 20 frames per second, corresponding to an acquisition time of approximately 5~s for 101 images. Subsets containing 1, 5, 10, 20, 50 or 101 images per curvature state were used to evaluate the influence of the available data volume on reconstruction performance. As shown in Figure~\ref{fig:fig6}(a,b), the RMSE and MAE decrease under both all-random and curvature-random splitting as more images become available at each curvature state. The largest reduction occurs within the first 10--20 images, followed by progressively smaller improvements as the errors approach a plateau. Curvature-random splitting provides a more stringent evaluation because complete curvature states are withheld from the training set, thereby assessing the model's ability to predict previously unseen curvature values. It was therefore used as the primary evaluation strategy for quantitative performance comparison. With 101 images available at each curvature state, the 4CF achieves an RMSE of \SI{0.0218}{\per\meter} and an MAE of \SI{0.0146}{\per\meter}. In the repeated-test evaluation shown in Figure~\ref{fig:fig6}(c), increasing the number of images beyond approximately 50 produces only marginal error reduction, while the computation time continues to increase approximately linearly. For the 4CF repeated-test evaluation, let $N$ denote the number of images per curvature state and $T$ the total training and prediction time for one evaluation run. For GBRT, the fitted slope of $T$ versus $N$ was approximately 1.8~ms per unit increase in $N$.

The repeated-test predictions in Figure~\ref{fig:fig6}(d) closely follow the ideal relationship between measured and predicted curvature, while the larger deviations at several curvature states indicate shifts in the mapping between curvature and the output intensity distribution, reflecting reduced repeatability of the measured response at these states. Figure~\ref{fig:fig6}(e) presents the error distributions from one representative evaluation to illustrate the shapes and locations of the prediction errors under the three splitting strategies.

\begin{figure}[tbp]
    \centering
    \includegraphics[width=\linewidth]{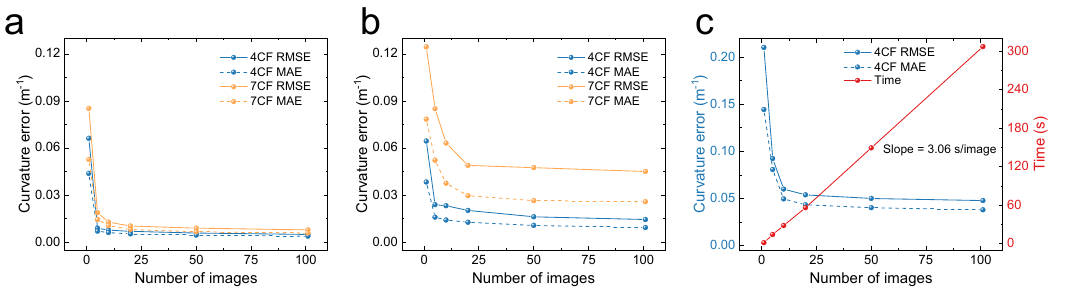}
    \caption{{MLP-based curvature reconstruction.}
        \textbf{a}, RMSE and MAE versus the number of images per curvature state for the 4CF and 7CF under all-random splitting.
        \textbf{b}, Corresponding errors under curvature-random splitting.
        \textbf{c}, RMSE, MAE, and total training and prediction time for the 4CF in the repeated-test evaluation. All results are averaged over 200 evaluations.}
    \label{fig:fig7}
\end{figure}

Figure~\ref{fig:fig7} presents the MLP reconstruction results using the same data-splitting strategies. The prediction errors also decrease as more images are included at each curvature state, as shown in Figure~\ref{fig:fig7}(a,b). Under curvature-random splitting with 101 images, the 4CF achieves an RMSE of \SI{0.014}{\per\meter} and an MAE of \SI{0.009}{\per\meter}. These errors are lower than those obtained using GBRT. Under all-random and repeated-test evaluations, the MLP generally produces larger errors than GBRT and requires substantially longer computation time. The corresponding slope for the MLP was approximately 3.06~s per unit increase in $N$, as shown in Figure~\ref{fig:fig7}(c). Among the tested implementations, GBRT provides a more favorable overall balance between prediction accuracy and computation time, while the MLP shows a potential advantage for prediction at unseen curvature states.

\begin{figure}[!htb]
    \centering
    \includegraphics{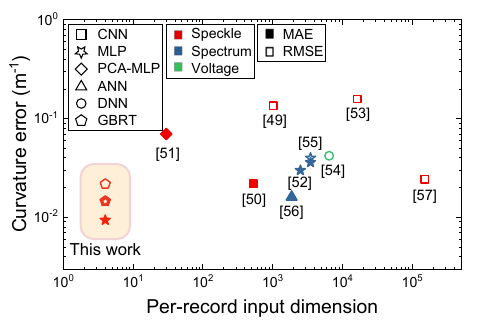}
    \caption{Comparison of input feature dimension and reported curvature errors for representative machine-learning-assisted fiber sensors. The highlighted 4CF results use four features per image for both MLP and GBRT under curvature-random splitting. Input dimension counts scalar model inputs after preprocessing. Filled and open symbols denote MAE and RMSE, respectively; reference numbers identify the literature sources. PCA-MLP denotes principal component analysis followed by MLP regression. CNN, convolutional neural network; MLP, multilayer perceptron; ANN, artificial neural network; DNN, deep neural network; GBRT, gradient-boosted regression trees.}
    \label{fig:fig8}
\end{figure}

Figure~\ref{fig:fig8} compares the per-record input dimension and reported curvature errors of representative machine-learning-assisted fiber sensors~\cite{Li2022OLT,Li2022SJ,Gao2023OE,Huang2023OL, Bender2023SJ,Pamukti2024JOSAB,Niu2025PTL,Ren2025SJ, Shen2025AOS}. Here, the input dimension denotes the number of scalar features supplied to the model after the study-specific preprocessing of an acquired image, spectrum or voltage trace. It does not represent the total training-set size, raw acquisition volume or computation time. The surveyed schemes use tens to approximately $1.5\times10^5$ input features per record, whereas the RC-MCF readout uses four or seven intensity values per image for the 4CF and 7CF, respectively, corresponding to reductions ranging from several-fold to more than four orders of magnitude across the surveyed schemes. For the 4CF, the MLP and GBRT achieve RMSE values of 0.014 and 0.022~m$^{-1}$, respectively, within the error range reported for the surveyed methods. Differences in curvature range, sampling, preprocessing and evaluation protocols make this a descriptive comparison of input feature count and reported error scale rather than a controlled performance ranking.

\section{Conclusion}

This study establishes random inter-core coupling as a reproducible physical encoding mechanism under a fixed fiber configuration. In conventional MCF transmission, coupling is commonly treated as crosstalk or a source of uncertainty. In the demonstrated sensing configuration, however, the coupled state remains stable in the absence of intentional deformation while responding sensitively to bending-induced changes in the relative propagation constants of the cores. The fixed multicore geometry further reduces the model input to one normalized intensity value per detected core, extracted from each camera image, allowing the sensing state to be represented by only four or seven normalized core-intensity features. Compared with conventional speckle-based schemes that rely on dense pixel sampling or large extracted feature sets, the proposed RC-MCF approach therefore combines perturbation-sensitive optical encoding with a compact and simplified readout, enabling lightweight regression with substantially reduced input dimensionality. Under curvature-random splitting, the multilayer perceptron achieved an RMSE of \SI{0.014}{\per\meter} and an MAE of \SI{0.009}{\per\meter}
when predicting held-out curvature states in the 4CF dataset.

Building on these results, future work will explore extending RC-MCF sensing toward more complex deformation measurements, including bending direction, deformation location, vectorial bending, and compound deformation states. An important consideration for such extensions is that remounting or rearranging the fiber can alter the initial inter-core optical power distribution and the subsequent deformation-dependent intensity evolution, potentially requiring recalibration or model retraining. To address this challenge, future investigations will focus on optimizing fiber structures and inter-core coupling characteristics to balance sensing sensitivity and robustness against configuration changes. In parallel, the current camera-based laboratory readout could be replaced by multicore-aligned photodiode arrays, multichannel power meters, or integrated optical receivers to increase acquisition speed and reduce data-processing demands. These developments, combined with lightweight regression, may further improve the practicality of RC-MCF sensing and support future applications in real-time vectorial bending and complex deformation monitoring.

\section*{Acknowledgements}
This work was supported by the National Natural Science Foundation of China (Grant No.~62275150).

\section*{Conflict of Interest}
The authors declare no conflict of interest.

\section*{Data Availability Statement}
The data that support the findings of this study are available from the corresponding author upon reasonable request.


\begin{thebibliography}{99}

\bibitem{ploschner2015seeing}
M.~Pl{\"{o}}schner, T.~Tyc, and T.~{\v{C}}i{\v{z}}m{\'{a}}r, \textquotedblleft{}Seeing through chaos in multimode fibres,\textquotedblright{} \emph{Nature Photonics} 9, no.~8 (2015): 529--535. \url{https://doi.org/10.1038/nphoton.2015.112}.

\bibitem{tzang2018adaptive}
O.~Tzang, A.~M. Caravaca-Aguirre, K.~Wagner, and R.~Piestun, \textquotedblleft{}Adaptive wavefront shaping for controlling nonlinear multimode interactions in optical fibres,\textquotedblright{} \emph{Nature Photonics} 12, no.~6 (2018): 368--374. \url{https://doi.org/10.1038/s41566-018-0167-7}.

\bibitem{li2021compressively}
S.~Li, C.~Saunders, D.~J. Lum, et~al., \textquotedblleft{}Compressively sampling the optical transmission matrix of a multimode fibre,\textquotedblright{} \emph{Light: Science \& Applications} 10, no.~1 (2021): 88. \url{https://doi.org/10.1038/s41377-021-00514-9}.

\bibitem{kupianskyi2024untangling}
H.~Kupianskyi, S.~A.~R. Horsley, and D.~B. Phillips, \textquotedblleft{}All-optically untangling light propagation through multimode fibers,\textquotedblright{} \emph{Optica} 11, no.~1 (2024): 101--112. \url{https://doi.org/10.1364/optica.502144}.

\bibitem{han2023tailoring}
S.~Han, N.~Bender, and H.~Cao, \textquotedblleft{}Tailoring {3D} speckle statistics,\textquotedblright{} \emph{Physical Review Letters} 130, no.~9 (2023): 093802. \url{https://doi.org/10.1103/physrevlett.130.093802}.

\bibitem{rahmani2018multimode}
B.~Rahmani, D.~Loterie, G.~Konstantinou, D.~Psaltis, and C.~Moser, \textquotedblleft{}Multimode optical fiber transmission with a deep learning network,\textquotedblright{} \emph{Light: Science \& Applications} 7, no.~1 (2018): 69. \url{https://doi.org/10.1038/s41377-018-0074-1}.

\bibitem{stellinga2021time}
D.~Stellinga, D.~B. Phillips, S.~P. Mekhail, et~al., \textquotedblleft{}Time-of-flight {3D} imaging through multimode optical fibers,\textquotedblright{} \emph{Science} 374, no.~6573 (2021): 1395--1399. \url{https://doi.org/10.1126/science.abl3771}.

\bibitem{luo2022computational}
Y.~Luo, Y.~Zhao, J.~Li, et~al., \textquotedblleft{}Computational imaging without a computer: seeing through random diffusers at the speed of light,\textquotedblright{} \emph{eLight} 2, no.~1 (2022): 4. \url{https://doi.org/10.1186/s43593-022-00012-4}.

\bibitem{xu2024wgm}
B.~Xu, Y.~Wan, X.~Fan, and Z.~He, \textquotedblleft{}Whispering-gallery-mode barcode-based broadband sub-femtometer-resolution spectroscopy with an electro-optic frequency comb,\textquotedblright{} \emph{Advanced Photonics} 6, no.~1 (2024): 016006. \url{https://doi.org/10.1117/1.ap.6.1.016006}.

\bibitem{gao2025semantic}
Z.~Gao, T.~Jiang, M.~Zhang, H.~Wu, and M.~Tang, \textquotedblleft{}Optical semantic communication through multimode fiber: from symbol transmission to sentiment analysis,\textquotedblright{} \emph{Light: Science \& Applications} 14, no.~1 (2025): 60. \url{https://doi.org/10.1038/s41377-024-01726-5}.

\bibitem{leedumrongwatthanakun2020programmable}
S.~Leedumrongwatthanakun, L.~Innocenti, H.~Defienne, et~al., \textquotedblleft{}Programmable linear quantum networks with a multimode fibre,\textquotedblright{} \emph{Nature Photonics} 14, no.~3 (2020): 139--142. \url{https://doi.org/10.1038/s41566-019-0553-9}.

\bibitem{facchin2024intrinsic}
M.~Facchin, S.~N. Khan, K.~Dholakia, and G.~D. Bruce, \textquotedblleft{}Determining intrinsic sensitivity and the role of multiple scattering in speckle metrology,\textquotedblright{} \emph{Nature Reviews Physics} 6, no.~8 (2024): 500--508. \url{https://doi.org/10.1038/s42254-024-00735-y}.

\bibitem{chapalo2024review}
I.~Chapalo, A.~Stylianou, P.~M{\'{e}}gret, and A.~Theodosiou, \textquotedblleft{}Advances in optical fiber speckle sensing: a comprehensive review,\textquotedblright{} \emph{Photonics} 11, no.~4 (2024): 299. \url{https://doi.org/10.3390/photonics11040299}.

\bibitem{smith2022specklegram}
D.~L. Smith, L.~V. Nguyen, D.~J. Ottaway, et~al., \textquotedblleft{}Machine learning for sensing with a multimode exposed core fiber specklegram sensor,\textquotedblright{} \emph{Optics Express} 30, no.~7 (2022): 10443--10455. \url{https://doi.org/10.1364/OE.443932}.

\bibitem{zhou2018diffusing}
W.~Zhou, O.~Kholiqov, S.~P. Chong, and V.~J. Srinivasan, \textquotedblleft{}Highly parallel, interferometric diffusing wave spectroscopy for monitoring cerebral blood flow dynamics,\textquotedblright{} \emph{Optica} 5, no.~5 (2018): 518--527. \url{https://doi.org/10.1364/optica.5.000518}.

\bibitem{huang2018dna}
F.~Feng, W.~Chen, D.~Chen, W.~Lin, and S.-C. Chen, \textquotedblleft{}In-situ ultrasensitive label-free {DNA} hybridization detection using optical fiber specklegram,\textquotedblright{} \emph{Sensors and Actuators B: Chemical} 272 (2018): 160--165. \url{https://doi.org/10.1016/j.snb.2018.05.099}.

\bibitem{bielak2026polarization}
M.~Bielak, D.~Va{\v{s}}inka, and M.~Je{\v{z}}ek, \textquotedblleft{}All-fiber microsensor of polarization at single-photon level aided by deep learning,\textquotedblright{} \emph{Laser \& Photonics Reviews} 20, no.~3 (2026): e01775. \url{https://doi.org/10.1002/lpor.202501775}.

\bibitem{zhou2021functional}
W.~Zhou, O.~Kholiqov, J.~Zhu, et~al., \textquotedblleft{}Functional interferometric diffusing wave spectroscopy of the human brain,\textquotedblright{} \emph{Science Advances} 7, no.~20 (2021): eabe0150. \url{https://doi.org/10.1126/sciadv.abe0150}.

\bibitem{tao2026nanosecond}
Y.~Tao, Y.~Wan, Z.~Long, W.~Zhang, J.~Du, and Z.~He, \textquotedblleft{}Nanosecond-latency all-optical fiber sensing with in-sensor computing,\textquotedblright{} \emph{Light: Science \& Applications} 15, no.~1 (2026): 251. \url{https://doi.org/10.1038/s41377-026-02265-x}.

\bibitem{xiong2024polarimeter}
Y.~Xiong, T.~Jiang, H.~Wu, et~al., \textquotedblleft{}Multimode fiber speckle {Stokes} polarimeter,\textquotedblright{} \emph{Advanced Photonics Nexus} 3, no.~4 (2024): 046010. \url{https://doi.org/10.1117/1.apn.3.4.046010}.

\bibitem{newaz2023review}
A.~Newaz, M.~O. Faruque, R.~A. Mahmud, R.~H. Sagor, and M.~Z.~M. Khan, \textquotedblleft{}Machine-learning-enabled multimode fiber specklegram sensors: a review,\textquotedblright{} \emph{IEEE Sensors Journal} 23, no.~18 (2023): 20937--20950. \url{https://doi.org/10.1109/jsen.2023.3298169}.

\bibitem{xia2024nonlinear}
F.~Xia, K.~Kim, Y.~Eliezer, et~al., \textquotedblleft{}Nonlinear optical encoding enabled by recurrent linear scattering,\textquotedblright{} \emph{Nature Photonics} 18, no.~10 (2024): 1067--1075. \url{https://doi.org/10.1038/s41566-024-01493-0}.

\bibitem{li2024ultrashort}
G.~Li, Y.~Liu, L.~Pang, H.~Yuan, and M.~Wang, \textquotedblleft{}A novel structure with ultra short multimode fiber for fiber specklegram sensor and its application in multi-bending sensing. {Proof} and concept,\textquotedblright{} \emph{Journal of Lightwave Technology} 42, no.~24 (2024): 8966--8974. \url{https://doi.org/10.1109/JLT.2024.3439347}.

\bibitem{wei2021limited}
M.~Wei, G.~Tang, J.~Liu, et~al., \textquotedblleft{}Neural network based perturbation-location fiber specklegram sensing system towards applications with limited number of training samples,\textquotedblright{} \emph{Journal of Lightwave Technology} 39, no.~19 (2021): 6315--6326. \url{https://doi.org/10.1109/JLT.2021.3098345}.

\bibitem{fu2024onn}
T.~Fu, J.~Zhang, R.~Sun, et~al., \textquotedblleft{}Optical neural networks: progress and challenges,\textquotedblright{} \emph{Light: Science \& Applications} 13, no.~1 (2024): 263. \url{https://doi.org/10.1038/s41377-024-01590-3}.

\bibitem{vandenhout2025pinnacle}
M.~van~den Hout, R.~S. Lu{\'{i}}s, B.~J. Puttnam, et~al., \textquotedblleft{}Reaching the pinnacle of high-capacity optical transmission using a standard cladding diameter coupled-core multi-core fiber,\textquotedblright{} \emph{Nature Communications} 16, no.~1 (2025): 3833. \url{https://doi.org/10.1038/s41467-025-59037-1}.

\bibitem{richardson2013sdm}
D.~J. Richardson, J.~M. Fini, and L.~E. Nelson, \textquotedblleft{}Space-division multiplexing in optical fibres,\textquotedblright{} \emph{Nature Photonics} 7, no.~5 (2013): 354--362. \url{https://doi.org/10.1038/nphoton.2013.94}.

\bibitem{sun2022qpi}
J.~Sun, J.~Wu, S.~Wu, et~al., \textquotedblleft{}Quantitative phase imaging through an ultra-thin lensless fiber endoscope,\textquotedblright{} \emph{Light: Science \& Applications} 11, no.~1 (2022): 204. \url{https://doi.org/10.1038/s41377-022-00898-2}.

\bibitem{zolnacz2024endoscopic}
K.~Zolnacz, R.~Stephan, J.~Dremel, et~al., \textquotedblleft{}Multicore fiber with thermally expanded cores for increased collection efficiency in endoscopic imaging,\textquotedblright{} \emph{Light: Advanced Manufacturing} 5, no.~4 (2024): 580--587. \url{https://doi.org/10.37188/lam.2024.049}.

\bibitem{klenke2018laser}
A.~Klenke, M.~M{\"{u}}ller, H.~Stark, et~al., \textquotedblleft{}Coherently combined 16-channel multicore fiber laser system,\textquotedblright{} \emph{Optics Letters} 43, no.~7 (2018): 1519--1522. \url{https://doi.org/10.1364/ol.43.001519}.

\bibitem{li2026hybrid}
K.~Li, G.~Yan, K.~Wang, et~al., \textquotedblleft{}Harnessing diverse hybrid integration for bridging trans-scale multi-dimensional fiber-chip data transmission and processing,\textquotedblright{} \emph{Light: Science \& Applications} 15, no.~1 (2026): 167. \url{https://doi.org/10.1038/s41377-026-02194-9}.

\bibitem{zahidy2024qkd}
M.~Zahidy, D.~Ribezzo, C.~De~Lazzari, et~al., \textquotedblleft{}Practical high-dimensional quantum key distribution protocol over deployed multicore fiber,\textquotedblright{} \emph{Nature Communications} 15, no.~1 (2024): 1651. \url{https://doi.org/10.1038/s41467-024-45876-x}.

\bibitem{caucheteur2022sensors}
C.~Caucheteur, J.~Villatoro, F.~Liu, M.~Loyez, T.~Guo, and J.~Albert, \textquotedblleft{}Mode-division and spatial-division optical fiber sensors,\textquotedblright{} \emph{Advances in Optics and Photonics} 14, no.~1 (2022): 1--86. \url{https://doi.org/10.1364/aop.444261}.

\bibitem{wolf2022fbg}
A.~Wolf, A.~Dostovalov, K.~Bronnikov, M.~Skvortsov, S.~Wabnitz, and S.~Babin, \textquotedblleft{}Advances in femtosecond laser direct writing of fiber {Bragg} gratings in multicore fibers: technology, sensor and laser applications,\textquotedblright{} \emph{Opto-Electronic Advances} 5, no.~4 (2022): 210055. \url{https://doi.org/10.29026/oea.2022.210055}.

\bibitem{matsui2022wcmcf}
T.~Matsui, P.~L. Pondillo, and K.~Nakajima, \textquotedblleft{}Weakly coupled multicore fiber technology, deployment, and systems,\textquotedblright{} \emph{Proceedings of the IEEE} 110, no.~11 (2022): 1772--1785. \url{https://doi.org/10.1109/JPROC.2022.3202812}.

\bibitem{diamandi2017crosstalk}
H.~H. Diamandi, Y.~London, and A.~Zadok, \textquotedblleft{}Opto-mechanical inter-core cross-talk in multi-core fibers,\textquotedblright{} \emph{Optica} 4, no.~3 (2017): 289--297. \url{https://doi.org/10.1364/optica.4.000289}.

\bibitem{soeiro2018bending}
R.~O.~J. Soeiro, T.~M.~F. Alves, and A.~V.~T. Cartaxo, \textquotedblleft{}Impact of longitudinal variation of the coupling coefficient due to bending and twisting on inter-core crosstalk in weakly-coupled {MCFs},\textquotedblright{} \emph{Journal of Lightwave Technology} 36, no.~18 (2018): 3898--3911. \url{https://doi.org/10.1109/JLT.2018.2852953}.

\bibitem{sakamoto2024nonlinear}
T.~Sakamoto, R.~Imada, and K.~Nakajima, \textquotedblleft{}Characterization of effective nonlinear coefficient in randomly coupled multi-core fiber,\textquotedblright{} \emph{Journal of Lightwave Technology} 42, no.~13 (2024): 4659--4665. \url{https://doi.org/10.1109/jlt.2024.3374766}.

\bibitem{xu2026lpr}
M.~Xu, L.~Ma, T.~Cheng, F.~Yu, and Z.~He, \textquotedblleft{}Randomly-coupled multicore fibers for ultra-sensitive curvature sensing,\textquotedblright{} \emph{Laser \& Photonics Reviews} 20, no.~9 (2026): e02379. \url{https://doi.org/10.1002/lpor.202502379}.

\bibitem{hu2025lstm}
J.~Hu, S.~Zhang, M.~Cai, et~al., \textquotedblleft{}{LSTM}-assisted optical fiber interferometric sensing: breaking the limitation of free spectral range,\textquotedblright{} \emph{Light: Science \& Applications} 14, no.~1 (2025): 392. \url{https://doi.org/10.1038/s41377-025-02008-4}.

\bibitem{li2025tactile}
L.~Li, F.~Yang, Q.~Ma, et~al., \textquotedblleft{}Integrating an optical fiber sensor and artificial intelligence for enhanced tactile sensing in human-computer interaction,\textquotedblright{} \emph{Optica} 12, no.~2 (2025): 263--273. \url{https://doi.org/10.1364/OPTICA.553706}.

\bibitem{arrizabalaga2020vector}
O.~Arrizabalaga, Q.~Sun, M.~Beresna, et~al., \textquotedblleft{}High-performance vector bending and orientation distinguishing curvature sensor based on asymmetric coupled multi-core fibre,\textquotedblright{} \emph{Scientific Reports} 10, no.~1 (2020): 14058. \url{https://doi.org/10.1038/s41598-020-70999-8}.

\bibitem{xiong2024corepitch}
J.~Xiong, L.~Ma, Y.~Chang, et~al., \textquotedblleft{}Study on characteristics of coupled-core four-core fibers with different core pitches,\textquotedblright{} \emph{IEEE Photonics Journal} 16, no.~1 (2024): 1--7. \url{https://doi.org/10.1109/jphot.2023.3341415}.

\bibitem{bentejac2021comparative}
C.~Bent{\'{e}}jac, A.~Cs{\"{o}}rg{\H{o}}, and G.~Mart{\'{i}}nez-Mu{\~{n}}oz, \textquotedblleft{}A comparative analysis of gradient boosting algorithms,\textquotedblright{} \emph{Artificial Intelligence Review} 54, no.~3 (2021): 1937--1967. \url{https://doi.org/10.1007/s10462-020-09896-5}.

\bibitem{hollmann2025accurate}
N.~Hollmann, S.~M{\"{u}}ller, L.~Purucker, et~al., \textquotedblleft{}Accurate predictions on small data with a tabular foundation model,\textquotedblright{} \emph{Nature} 637, no.~8045 (2025): 319--326. \url{https://doi.org/10.1038/s41586-024-08328-6}.

\bibitem{rumelhart1986backprop}
D.~E. Rumelhart, G.~E. Hinton, and R.~J. Williams, \textquotedblleft{}Learning representations by back-propagating errors,\textquotedblright{} \emph{Nature} 323, no.~6088 (1986): 533--536. \url{https://doi.org/10.1038/323533a0}.

\bibitem{smith2024complex}
D.~L. Smith, L.~V. Nguyen, M.~I. Reja, et~al., \textquotedblleft{}Harnessing the power of complex light propagation in multimode fibers for spatially resolved sensing,\textquotedblright{} \emph{Photonics Research} 12, no.~3 (2024): 411--422. \url{https://doi.org/10.1364/prj.507542}.

\bibitem{Kingma2015Adam}
D.~P. Kingma and J.~Ba, \textquotedblleft{}{Adam}: a method for stochastic optimization,\textquotedblright{} in \emph{{3rd International Conference on Learning Representations (ICLR)}} (2015). \url{https://arxiv.org/abs/1412.6980}.

\bibitem{Li2022OLT}
G.~Li, Y.~Liu, Q.~Qin, X.~Zou, M.~Wang, and F.~Yan, \textquotedblleft{}Deep learning based optical curvature sensor through specklegram detection of multimode fiber,\textquotedblright{} \emph{Optics \& Laser Technology} 149 (2022): 107873. \url{https://doi.org/10.1016/j.optlastec.2022.107873}.

\bibitem{Li2022SJ}
G.~Li, Y.~Liu, Q.~Qin, X.~Zou, M.~Wang, and W.~Ren, \textquotedblleft{}Feature extraction enabled deep learning from specklegram for optical fiber curvature sensing,\textquotedblright{} \emph{IEEE Sensors Journal} 22, no.~16 (2022): 15974--15984. \url{https://doi.org/10.1109/JSEN.2022.3188694}.

\bibitem{Gao2023OE}
H.~Gao and H.~Hu, \textquotedblleft{}Spatially-resolved bending recognition based on a learning-empowered fiber specklegram sensor,\textquotedblright{} \emph{Optics Express} 31, no.~5 (2023): 7671--7683. \url{https://doi.org/10.1364/OE.482953}.

\bibitem{Huang2023OL}
B.~Huang, X.~Sheng, J.~Cao, et~al., \textquotedblleft{}Machine-learning-assisted omnidirectional bending sensor based on a cascaded asymmetric dual-core {PCF} sensor,\textquotedblright{} \emph{Optics Letters} 48, no.~19 (2023): 4929--4932. \url{https://doi.org/10.1364/OL.497509}.

\bibitem{Bender2023SJ}
D.~Bender, U.~{\c{C}}ak{\i}r, and E.~Y{\"{u}}ce, \textquotedblleft{}Deep learning-based fiber bending recognition for sensor applications,\textquotedblright{} \emph{IEEE Sensors Journal} 23, no.~7 (2023): 6956--6962. \url{https://doi.org/10.1109/JSEN.2023.3249049}.

\bibitem{Pamukti2024JOSAB}
B.~Pamukti, M.~F.~F. Pradipta, S.-K. Liaw, F.-L. Yang, and Y.-M. Yang, \textquotedblleft{}Deep learning method for optical fiber curvature measurements based on time series data,\textquotedblright{} \emph{Journal of the Optical Society of America B} 41, no.~5 (2024): 1207--1216. \url{https://doi.org/10.1364/JOSAB.517639}.

\bibitem{Niu2025PTL}
C.~Niu, Y.~Wang, Y.~Kou, et~al., \textquotedblleft{}Machine learning assisted high precision vector bending sensor based on remodulate {LPFG},\textquotedblright{} \emph{IEEE Photonics Technology Letters} 37, no.~2 (2025): 113--116. \url{https://doi.org/10.1109/LPT.2024.3516133}.

\bibitem{Ren2025SJ}
L.~Ren, L.~Li, Q.~Yu, B.~Han, and Y.-N. Zhang, \textquotedblleft{}{SVM}-{ANN}-assisted asymmetrical optical fiber {MZI} for accurate vector bending sensing,\textquotedblright{} \emph{IEEE Sensors Journal} 25, no.~16 (2025): 31500--31507. \url{https://doi.org/10.1109/JSEN.2025.3587630}.

\bibitem{Shen2025AOS}
Z.~Shen, L.~Cai, G.~Yang, and S.~Liu, \textquotedblleft{}Curvature measurement method based on {VggNet} 16 and specialty optical fiber speckle,\textquotedblright{} \emph{Acta Optica Sinica} 45, no.~15 (2025): 1506001. \url{https://doi.org/10.3788/AOS250693}.

\end{thebibliography}
\end{document}